\documentclass[prl,aps,twocolumn,nofootinbib]{revtex4}
\usepackage{epsf}

\def\half{{1\over 2}}
\def\x{{\bf x}}
\def\n{{\bf n}}
\def\A{{\bf A}}
\def\btheta{\mbox{\boldmath $\theta$}}
\def\cap{{\cal C}}
\def\ind{{\cal L}}
\def\M{{\cal M}}
\def\be{\begin{equation}}
\def\ee{\end{equation}}
\def\ba{\begin{eqnarray*}}
\def\ea{\end{eqnarray*}}
\begin{document}
\title{Electron as soliton: Nonlinear theory of dielectric polarization}
\author{S. Khlebnikov}
\affiliation{Department of Physics, Purdue University, West Lafayette,
Indiana 47907, USA}

\begin{abstract}
We describe a class of theories of dielectric polarization and a species of
solitons in these theories. The solitons, made entirely out of 
the polarization field, have quantized values of the electric charge and can
be interpreted as electrons and holes. A soliton-antisoliton pair
is an exciton. We present numerical evidence that 
the elementary soliton is stable.
\end{abstract}

\maketitle

Absorption of light by dielectrics is one of the oldest problems in 
quantum theory.
Because a typical frequency of visible light is much larger than the maximal
frequency of phonons, absorption of even one photon must result in many
phonons. Frenkel's idea \cite{Frenkel} was that the immense suppression, naively
expected for such a process, can be overcome if there is an intermediary---an
electronic excitation of the solid (an exciton).
An alternative to this point of view would be to imagine that phonons produced
by light absorption initially form
a highly coherent, nearly classical state. Then, the effective
nonlinearity is enhanced by the large amplitude of the phonon field in the final
state, resulting in a sizable absorption rate.

In the present Letter, we would like to argue that these two points of view
are, in fact, complementary to each
other. Namely, if the phonon field is treated as a nonlinear system, there are
solitons of this field that are naturally interpreted as electrons and holes,
and a soliton-antisoliton pair can be interpreted as an exciton.

We wish to stress from the outset that our results are quite different from the
familiar polaron picture. In the latter case, there are two fields,
which describe electrons and phonons separately, and the polaron is a bound
state involving both. In our case, there is no separate electron field---electrons
emerge as nonlinear excitations of phonons, somewhat similarly to how a baryon emerges
as a nonlinear excitation of pions in Skyrme's model \cite{Skyrme}. 
An even closer similarity
is with the solitons of the sine-Gordon (SG) model, which can be viewed 
\cite{Coleman,Mandelstam} as fermions of the massive Thirring model.

Neither of these similarities is complete, however.
Both in the SG case and in Skyrme's model, solitons exists directly 
in the continuum limit,
while in our case, at least in the present version of the theory,
they can only live on a lattice.

One reason to pursue the picture of electrons as solitons is that it allows one
to think about various processes of interest in nanoscale dielectrics entirely 
in classical terms. For example, one may want to know 
if a highly excited excitonic state
relaxes to the lowest-energy exciton via carrier multiplication (i.e., production
of more excitons) or via emission of multiple phonons. By viewing electrons as 
solitons of the phonon field, 
we obtain a classical approximation, which can be used to simulate this process
on a computer.

Quantum physics of these solitons, in particular, the question of their spin and
statistics, also promises to be interesting. In the present Letter, however, we limit
ourselves to problems of the classical theory: arguing the existence of solitons and
describing their simplest properties. While a plausible existence argument can be
given analytically, as further evidence we also present results 
of a numerical study.

Consider then the field $\theta_n(\x,t)$ of an optical phonon. We will
consider in parallel spatial dimensions $d=2$ and 3 (planar and bulk dielectrics,
respectively). The number of components of $\theta_n$, i.e., the number of values
that $n$ takes, is equal to $d$. We assume that $\theta_n$ is defined on
the faces (for $d=3$) or edges (for $d=2$) of a rectangular lattice with spacings
$a_n$. Up to a constant factor, $\theta_n$ is the component of dielectric polarization
in the $n$-th direction. For notational 
simplicity, we will often use continuum notation for lattice derivatives.
Thus, for example, the electric charge density (in suitable units) will be written
as a divergence,
\be
\rho(\x, t) = \nabla \cdot \btheta (\x,t) \equiv \partial_n \theta_n (\x,t) \; ,
\label{rho}
\ee
but this should be read to mean
\be
\rho(\tilde{\x}, t) = 
\sum_{n=1}^d \frac{1}{a_n} [ \theta_n(\x + \n, t) - \theta_n(\x, t) ] \; ,
\label{lat}
\ee
where $\n$ is the primitive vector in the $n$-th direction. Note that, since $\theta_n$
live on the faces (or edges), $\rho$ naturally lives in lattice cells 
(i.e., on the lattice dual to the original). Hence the tilde in eq. (\ref{lat}).

The form of the Lagrangian density that governs dynamics of $\theta_n$ should 
ultimately be based on the properties of the material. For our present purposes,
it will be sufficient to use the simplest form, which we write as the first few
terms of a derivative expansion:
\be
L = - V(\btheta) - \frac{1}{2\cap} (\nabla \cdot \btheta)^2
+ \frac{\ind}{2} (\partial_t \btheta)^2 
+ \half \M_{mn} (\partial_t  \partial_m \theta_n )^2 \; ,
\label{L}
\ee
where $\cap > 0$, $\ind$, and $\M_{mn}$ are constants. In what follows, we limit ourselves
to searching for static solutions of the theory (\ref{L}), and for these the terms
containing time derivatives do not matter. 
They would be crucial, however, if we were to study the phonon spectrum.

The potential $V$ contains no derivatives of $\theta_n$
and describes the dielectric response of the medium to 
uniform static fields. The only feature of it essential here is a certain 
periodicity, so we use the simple cosine form (which in one dimension would be 
the potential of the SG model):
\be
V(\btheta) = \mu^2 \sum_n \frac{1}{\sigma_n^2} [1- \cos( \sigma_n \theta_n) ] \; .
\label{V}
\ee
In (\ref{V}), $\mu^2$ is a parameter, and $\sigma_n$ is the area of the faces 
(for $d=2$, length of the edges) that are orthogonal to the $n$-th direction. 
At small
$\theta_n$, $V \approx \half \mu^2 \theta_n^2$, which allows one to relate $\mu^2$ 
to the bulk dielectric constant $\epsilon$: $\mu^2 =e^2 /\pi (\epsilon -1)$,
where $e^2=14.4~\mbox{eV~\AA}$ is the electron charge squared.

The second and third terms on the right-hand side of (\ref{L}) describe the
capacitive and inductive effects due to short-scale charge separation.
In particular, 
\be
l_p = 1/\mu \sqrt{\cap} 
\label{lp}
\ee
is a characteristic length for spatial variations of 
the polarization charge. For a material with a high degree of spatial symmetry,
where we expect the expression (\ref{L}) to apply,
$\cap$, together with $\ind$ and $\M_{mn}$, can in principle 
be determined by fitting the phonon spectrum. 

Note that if $\mu$ is small (i.e., $\epsilon$ is large), 
as in many semiconductors, $l_p$ can be much larger than the atomic scale.
In this case, 
in nanocrystals of sufficiently small size,
effects of spatial dispersion may be quite significant. On the experimental side,
deviations of 
measured excitonic spectra from predictions of the
``envelope'' theory (in which electrons are considered point-like) have been reported
for PbSe crystals with diameters smaller than 7 nm \cite{Lipovskii&al}.

The short-scale effects encoded in the values of $\cap$, $\ind$, and $\M_{mn}$
should be
distinguished from effects due to macroscopic electromagnetic (e.-m.) fields
that can exist in a dielectric. To take those fields into
account, we would need to add to (\ref{L}) the term
\be
L' = \frac{|e|}{2\pi} \left[
A_0 \nabla \cdot \btheta + \frac{1}{c} \A \cdot \partial_t \btheta \right] \; ,
\label{L'}
\ee
where $A_0$ and $\A$ are the scalar and vector potentials, $e$ is the electron charge, 
and $c$ is the speed of light in vacuum. In $d=3$, this term also contributes
to the phonon frequency. For the moment, we are not interested
in macroscopic e.-m. fields of solitons, so in what follows we omit the Lagrangian 
(\ref{L'}).  Effects of impurities
are also readily incorporated into the theory, by adding a term 
$v(\x) \nabla \cdot \btheta(\x,t)$, where $v(\x)$ is the impurity potential. For now,
these will be neglected as well.

An important property of eq. (\ref{L}) is the absence of ``transverse capacitances'',
i.e., terms of the form $(\partial_m \theta_n)^2$ with $m \neq n$. As we will see,
such terms lead to a linear (confining) potential between solitons. Since
our goal is to describe a material in which electrons and holes can be separated away
to large distances, these terms are in fact forbidden. On the other hand, we could
add terms of the form $[\partial_m \sin (\sigma_n\theta_n)]^2$. 
At small $\theta_n$, 
these are indistinguishable from the forbidden terms, but at larger $\theta_n$ they 
display the periodicity property that causes the tension of the ``string'' connecting
solitons to vanish. If these terms are relatively small, then adding them to 
(\ref{L}) will only deform our solution a little, and for the present work we have 
left them out. In applications to specific materials, they can be readily included.

\begin{figure}[t]
\leavevmode\epsfxsize=2in \epsfbox{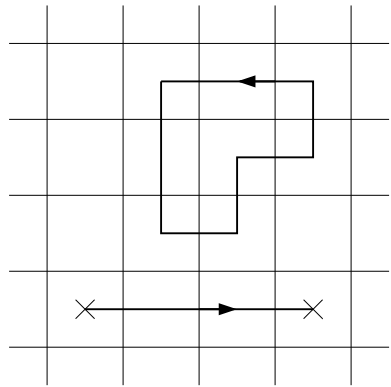}
\caption{The original lattice (thin lines) and closed and open strings on the dual 
lattice.}
\label{fig:strings}
\end{figure}

The argument forbidding the ``transverse capacitan\-ces'' can be made more formal
by identifying a relevant symmetry. For definiteness, we will speak about the $d=2$
(planar) case; the argument for $d=3$ is entirely similar. 
Consider a directed closed contour on the dual lattice (see Fig. \ref{fig:strings}).
Imagine changing the values of
$\theta_n$ at all the edges (in $d=3$, faces) the contour crosses by 
$\pm 2\pi / \sigma_n$, the sign depending on whether the contour runs parallel
or opposite to the $n$-th axis.
This transformation will be referred to as adding a closed
$2\pi$ string. The symmetry in question is the requirement that under this 
transformation the energy does not charge.

An open $2\pi$ string, such as the one shown in the lower portion 
of the figure, does cost energy, i.e., it is an excitation. If the symmetry 
with respect to adding closed strings is in place, however, the string tension 
is zero, and the energy of an open string is accumulated only near the ends. 
Since a $2\pi$ string carries  
electric flux of $2\pi$, the
ends of an open string are charged, with quantized charges equal to $\pm 2\pi$,
so this excitation can be 
interpreted as an electron-hole pair.

In the one-dimensional SG model, this way of creating excitations leads
to a soliton-antisoliton pair that is nearly stable (the more so the larger is 
the distance between the solitons). Indeed, imagine starting from the state with 
$\theta_n = 0$ (in one dimension, $\theta_n$
has only one component) and then changing $\theta_n$ by $2\pi$ at all sites of 
a chosen 
segment. If one then ``cools'' down the system, to reach a low-energy state,
the field will become smooth at the ends of
the segment but will remain equal to $2\pi$ in the middle: due to the high potential
barrier, it cannot unwind back to zero there. Similarly, in our case, we also
expect formation of a stable soliton pair (and, as we discuss shortly, this 
has been confirmed in numerical simulations).

If, however, we have allowed the ``transverse capacitances'', the symmetry 
with respect to adding closed $2\pi$ strings would be broken. 
For an open string, energy would then have come
not only from the ends but from the entire length and, for a large
length, would be proportional to it. As a result, a pair of well separated soliton
and antisoliton would not be an approximate solution: the linear potential would
pull the soliton and antisoliton towards each other until they annihilate. 
(This has also been confirmed numerically.) 

The symmetry with respect to adding closed $2\pi$ strings is local, in the sense 
that the number of independent transformations scales linearly with the volume of 
the system. Indeed, any closed
loop can be viewed as a superposition of elementary closed
loops, each encircling just a single site (in $d=2$) or edge (in $d=3$)
of the original lattice. With the loop's directionality taken into account,
there are twice as many ways to add an elementary loop
as there are sites or edges, respectively.

The local character of the symmetry suggests that the configuration space of our
theory is not the space of all possible $\theta_n(\x)$, but
the space of equivalence classes with respect to adding an arbitrary number of
closed $2\pi$ strings, i.e., states that differ only by the number of closed
$2\pi$ strings should be considered as one state. 
The role of this equivalence relation will become clearer shortly, 
when we consider the limit in which our soliton turns into a vortex or 
a magnetic monopole. However, this relation is not 
essential in the method we have used for finding solitons.

The presence of a string, similar to Dirac's string, suggests that our solution
is related to the magnetic monopole (or the vortex in $d=2$). Indeed, it is
a generalization of those. The monopole is obtained in the limit $\cap \to 0$.
Let us first consider this as a formal limit, i.e., as the requirement that
$\nabla \cdot \btheta = 0$. Then, $\btheta$ is the curl of some other field,
$\phi$. In $d=2$, this field lives on the sites of the original lattice and has
only one component, so that
\be
\theta_n = \epsilon_{nm} \partial_m \phi
\label{phi}
\ee
($\epsilon_{nm}$ is the antisymmetric unit tensor); in $d=3$, $\phi$ lives on 
the edges.
Changing $\phi$ by $2\pi$ at a single site creates, in terms of $\btheta$, an 
elementary closed $2\pi$ string encircling that site. So, imposing the 
equivalence relation as above amounts to requiring that $\phi$ is
an angular variable with period $2\pi$. 
Substituting eq. (\ref{phi}) into eq. (\ref{V}), we see that
the potential $V(\btheta)$ becomes the energy density of the 2-dimensional XY model, 
which has well-known static solutions---vortices. In $d=3$, a similar argument leads 
to compact electrodynamics, whose solutions are magnetic monopoles \cite{Polyakov}.
(They are
``magnetic'' with respect to the gauge field that replaces the scalar $\phi$ in
the curl condition, not with respect to ordinary electromagnetism.)

A less formal way to view the $\cap \to 0$ limit is to note that in this case
the screening length $l_p$, eq. (\ref{lp}), goes to infinity. Since the total 
charge of an elementary soliton is fixed (and equal to $2\pi$), the charge 
{\em density} $\nabla \cdot \btheta$
goes to zero. Thus, if a soliton and an antisoliton are separated by 
a distance much smaller than $l_p$, i.e., sit well within each other's polarization 
cloud, we expect them to
interact as if they were monopoles (or vortices) of the respective $\cap = 0$
theories.

As well known, monopoles and vortices have long-range interactions, which are 
important in various problems of statistical physics \cite{Polyakov}. From
our present standpoint, these interactions are a feature of the
$l_p \to \infty$ limit. For a finite $l_p$, the interaction range is of order $l_p$
(provided, of course, that the e.-m. interactions (\ref{L'}) have been switched off).

We now turn to numerical results. To search for static
solutions of the theory (\ref{L}), we drop the terms with time derivatives
and numerically solve the relaxation equation:
\be
\frac{\partial \btheta}{\partial \tau} = \nabla (\nabla \cdot \btheta) 
- \frac{1}{l_p^2} \frac{\partial V}{\partial \btheta} \; ,
\label{relax}
\ee
where $\tau$ is a fictitious time. To avoid influence of boundaries, 
we have used periodic boundary conditions.

As the initial condition, we have used an open string, such as shown in fig. 1, 
with various values of the electric flux. The actual solutions
(endpoints of the relaxation process) are expected to have the
flux quantized in integral multiples of $2\pi$---these are the values of 
$\sigma_n \theta_n$ at which the potential (\ref{V}) has minima. It is interesting,
however, to start with an arbitrary value and see how the nearest quantized value
is approached. Indeed, some of the most illustrative results are obtained by
starting with $\sigma_n \theta_n$ close to a {\em maximum} of the potential.

Fig. 2 shows results from a 2-dimensional $64\times 32$ lattice for the case when the
initial string has a flux of $\sigma_1 \theta_1 = 3.02 \pi$; results from
a 3-dimensional $64\times 32^2$ lattice (for the same initial flux) are similar. 
The screening length ($l_p = 1$) in either case was
about 3 times the lattice spacing (which was the same in all directions). 
\begin{figure}[t]
\leavevmode\epsfxsize=3.25in \epsfbox{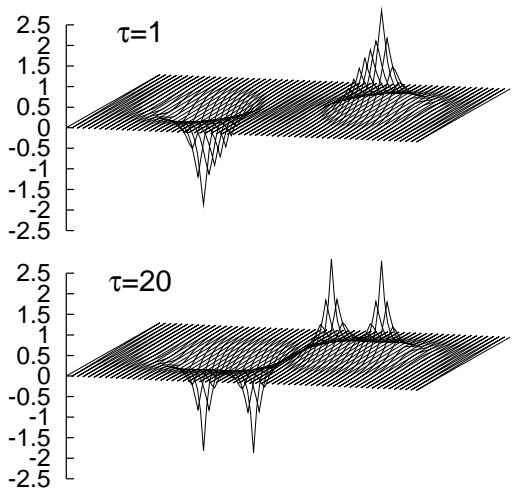}
\caption{Surface plots of the charge density towards the beginning (top) and end
(bottom) of the relaxation process.}
\label{fig:peaks}
\end{figure}

The initial flux
is just larger than $3\pi$, a value at which the potential has a maximum. So,
we expect the flux (and therefore also the charge at the string's end) to roll
to the nearest stable value, $\sigma_1 \theta_1 = 4\pi$, which is twice the minimal
soliton charge. The final value of the charge at each end is indeed close to
$\pm 4\pi$ (a small difference from $\pm 4\pi$ being attributable to finite-size 
effects), but we see that it is achieved by forming two solitons at each end,
rather than a single doubly-charged one. The final state was stable on 
the timescale of our simulation.

We have observed formation of two solitons instead of one
even when the initial string flux already was $\sigma_1 \theta_1 = 4\pi$.
We interpret this as an indication that the short-range 
interaction between solitons
of like charge is repulsive, and a compound soliton is unstable with respect to
decay into elementary ones. 

To summarize, we have described a class of theories that provide a unified 
description 
of phonons and excitons in a dielectric. Phonons are small fluctuations 
of a polarization field, and excitons are soliton-antisoliton pairs made of that same 
field. We have presented numerical evidence that the elementary soliton is stable and 
identified the symmetry (adding closed $2\pi$ strings) that underlies both this
stability and the quantization of the soliton charge.


\begin{thebibliography}{99}
\bibitem{Frenkel} J. Frenkel, Phys. Rev. 37, 17 (1931).
\bibitem{Skyrme} T. H. R. Skyrme, Proc. R. Soc. A 260, 127 (1961).
\bibitem{Coleman} S. Coleman, Phys. Rev. D 11, 2088 (1975).
\bibitem{Mandelstam} S. Mandelstam, Phys. Rev. D 11, 3026 (1975).
\bibitem{Lipovskii&al} A. Lipovskii {\em et al.}, Appl. Phys. Lett. 71,
3406 (1997).
\bibitem{Polyakov} For an introduction to vortices and monopoles on a lattice,
see A. M. Polyakov, {\em Gauge Fields and Strings} (Harwood, Chur, 1987).

\end{thebibliography}
\end{document}